\documentclass[12pt,thmsa]{article}
\usepackage{amsmath}
\usepackage{amsfonts}

\input{tcilatex}
\begin{document}

\title{Quantum field propagator for extended-objects in the microcanonical
ensemble and the S-matrix formulation.}
\author{Diego.J. Cirilo-Lombardo \\
{\small Bogoliubov Laboratory of Theoretical Physics}\\
{\small Joint Institute for Nuclear Research, 141980, Dubna, Russian
Federation.}\\
{\small e-mails: diego@thsun1.jinr.ru ; diego77jcl@yahoo.com}}
\maketitle

\begin{abstract}
Starting with the well-known Nambu-Goto action for an N-extended body system
the propagator in the microcanonical ensemble is explicitly computed. This
propagator is independent of the temperature and, in contrast with the
previous references, takes account on all the non-local effects produced by
the extended objects (e.g., strings) in interaction. The relation between
relativistic quantum field theories in the microcanonical approach and the
pure S-matrix formulation is stablished and analyzed.


PACS:
\end{abstract}

\section{\protect\bigskip \protect\bigskip Introduction}

Several studies into the behaviour of relativistic quantum field theories at
finite temperature have advanced considerably since the initial papers in
1974 [1-3]. In addition to the imaginary time formulation of Matsubara, two
more methods are available for formulating real time perturbation theory
with temperature dependent propagators: the functional integral approach
[5], and the operator approach of [6]. The calculations have all been
realized in the canonical ensemble, specified by volume $V$ \ held in
contact with a heat bath at a fixed temperature $T$. The partition function
and the scalar propagator were defined by [7] 
\begin{equation}
Z_{\beta }=\underset{n}{\sum }\exp (-\beta E_{n})
\end{equation}%
\begin{equation}
Z_{\beta }D_{\beta }(t,\mathbf{x})=-i\underset{n}{\sum }\exp (-\beta
E_{n})\left\langle n\right\vert T\left[ \varphi \left( t,\mathbf{x}\right)
\varphi \left( 0,0\right) \right] \left\vert n\right\rangle  \label{2}
\end{equation}%
where $\left\vert n\right\rangle $ label a complete set of energy
eigenstates and $\beta =1/T$, as usual. Since these propagators have
singularities whose locations depend on the temperature, some
inconsistencies appear: when the temperature is non zero the energy
differences that are independent of $T$ lead to singularities whose
locations are, apparently, $T$-dependent. To solve this puzzle, in [8] it
was shown that the natural context to study the same field theory is in the
microcanonical ensemble where the system is completely isolated at volume $V$%
, then the total energy of the system $E$ will remain constant. The
microcanonical partition function and propagator were naturally defined as 
\begin{equation}
\overline{Z}_{E}=\underset{n}{\sum }\delta (E_{n}-E)  \label{3}
\end{equation}%
\begin{equation}
\int_{0}^{E}dE^{\prime }\overline{Z}_{E-E^{\prime }}\overline{D}_{E^{\prime
}}(t,\mathbf{x})\equiv -i\underset{n}{\sum }\delta (E_{n}-E)\left\langle
n\right\vert T\left[ \varphi \left( t,\mathbf{x}\right) \varphi \left(
0,0\right) \right] \left\vert n\right\rangle  \label{4}
\end{equation}%
The mapping between the ensembles is realized by Laplace transform from $E$
to $\beta $. This distribution is more physically compelling in applying
quantum field theory to the early universe or to heavy-ion collisions since
the systems are completely isolated and are not certainly in contact with a
heat reservoir.

By the other hand, in ref.[19] the connection between dual amplitudes (i.e.,
interacting strings) and termodynamics properties of a strongly interacting
system was studied in the context of the canonical approach of ref.[18]. As
a result from this study, the authors of [19] showed that the strong duality
is loss in this strongly interacting system: the narrow resonances
aproximation does not reproduces the Regge model results.

In the this work, strongly motivated for above arguments, we give the next
step in studying relativistic quantum field theories in the context of the
microcanonical formulation : firstly, establish the relation with the
S-matrix formulation following a similar procedure as in refs.[12,18] in the
context of the canonical ensemble; and secondly, compute the microcanonical
propagator for a geometrical non-local and nonlinear action that is the
well-known Nambu-Goto action for a system of strings. The plan of this paper
is as follows: in Section 2, we describe the microcanonical formulation
briefly. In Section 3, the relation between the microcanonical approach in
relativistic quantum field theories and the axiomatic S-matrix formulation
is established. Finally, in Section 4, the microcanonical propagator for an
N-extended body system described by the Nambu-Goto action is successfully
performed and analyzed.

\section{\noindent Microcanonical formulation}

The true vacuum for a statistical system in thermal equilibrium can be
obtained writing the termal vacuum in terms of the density matrix $\widehat{%
\rho }$ 
\begin{equation}
\left\vert 0\left( \beta \right) \right\rangle =\widehat{\rho }\left( \beta ,%
\mathbf{H}\right) \left\vert \mathcal{J}\right\rangle ,  \label{5}
\end{equation}%
where 
\begin{equation}
\widehat{\rho }\left( \beta ,\mathbf{H}\right) =\frac{\rho \left( \beta ,%
\mathbf{H}\right) }{\left\langle \mathcal{J}\right\vert \rho \left( \beta ,%
\mathbf{H}\right) \left\vert \mathcal{J}\right\rangle }  \label{6}
\end{equation}%
\begin{equation}
\rho \left( \beta ,\mathbf{H}\right) =e^{-\beta \mathbf{H}}  \label{7}
\end{equation}%
\begin{equation}
\left\vert \mathcal{J}\right\rangle =\left[ \underset{k,m}{\prod }\underset{%
n_{k,m}}{\ \sum }\right] \underset{k,m}{\prod }\left\vert
n_{k,m}\right\rangle \otimes \left\vert \widetilde{n}_{k,m}\right\rangle
\label{8}
\end{equation}%
The trace of an observable operator is given by 
\begin{equation}
Tr\widehat{O}=\left\langle \mathcal{J}\right\vert \widehat{O}\left\vert 
\mathcal{J}\right\rangle  \label{9}
\end{equation}%
For example, the free field propagator can be determined from 
\begin{equation}
\Delta _{\beta }^{ab}=-i\left\langle \mathcal{J}\right\vert T\phi ^{a}\left(
x_{1}\right) \phi ^{b}\left( x_{2}\right) \widehat{\rho }\left\vert \mathcal{%
J}\right\rangle ,  \label{10}
\end{equation}%
where the superscripts on $\phi $ refer to the member of the termal doublet
being considered (for details see ref.[9]) 
\begin{equation*}
\phi ^{a}=\left( 
\begin{array}{l}
\phi \\ 
\widetilde{\phi }^{\dagger }%
\end{array}%
\right)
\end{equation*}%
As was shown in [9], the Fourier transform of $\Delta _{\beta }^{11}\left(
x_{1},x_{2}\right) $, that is the physical component, is equal to 
\begin{equation}
\Delta _{\beta }=\frac{1}{k^{2}-m^{2}+i\epsilon }-2\pi i\delta \left(
k^{2}-m^{2}\right) n_{\beta }\left( m,k\right)  \label{11}
\end{equation}%
where$\allowbreak $ in a canonical ensemble; for a thermal system at
temperature $\beta ^{-1}$ the number density is given by $n_{\beta }\left(
\omega \right) =\frac{1}{e^{\beta \omega }-1}$ .

If instead of treating the n-body system as objects in thermal equilibrium
at fixed temperature $T$ and corresponding vacuum $\left\vert \beta
_{H}\right\rangle $ we treat the system as having fixed energy $E$, we can
formally define the microcanonical vacuum as [10]: 
\begin{equation}
\left\vert E\right\rangle =\frac{1}{\Omega \left( E\right) }%
\int_{0}^{E}\Omega \left( E-E^{\prime }\right) L_{E-E^{\prime }}^{-1}\left[
\left\vert \beta _{H}\right\rangle \right] dE^{\prime },  \label{12}
\end{equation}%
where $L^{-1}$ is the inverse Laplace transform. Using this basis, the
physical correlation functions are expressed as 
\begin{equation}
G_{E}^{a_{1},...,a_{N}}\left( 1,2,....,N\right) =\left\langle \mathcal{J}%
\right\vert T\phi ^{a_{1}}\left( 1\right) ,.....,\phi ^{a_{N}}\left(
N\right) \left\vert E\right\rangle  \label{13}
\end{equation}%
The physical component of the propagator in the microcanonical field
formulation [9,10] is given by 
\begin{equation}
\Delta _{E}^{11}\left( k\right) =\frac{1}{k^{2}-m^{2}+i\epsilon }-2\pi
i\delta \left( k^{2}-m^{2}\right) n_{E}\left( m,k\right) ,  \label{14}
\end{equation}%
where the second term in the above expression corresponds to the statistical
(microcanonical) part and 
\begin{equation}
n_{E}\left( m,k\right) =\overset{\infty }{\underset{l=1}{\sum }}\frac{\Omega
\left( E-l\omega _{k}\left( m\right) \right) }{\Omega \left( E\right) }%
\theta \left( E-l\omega _{k}\right)  \label{15}
\end{equation}%
is the microcanonical number density, where $l,\omega _{k}$ and $\theta
\left( E-l\omega _{k}\right) $ are the mode number, the dispersion relation
and the step function, respectively. $\allowbreak $

\section{Axiomatic S-matrix formulation in QFT: microcanonical description}

The relation between the S matrix formulation in QFT and the statistical
operator in the canonical ensemble [11,12,18] can be easily extended to the
microcanonical description as follows. The propagator operator (we assume it
with its statistical part) 
\begin{equation}
\widehat{G}\left( E\right) =\frac{1}{E-\widehat{H}-i\epsilon }+G_{st},
\label{16}
\end{equation}%
namely, its imaginary part (where $G_{st}$ is the statistical part of the
full propagator) and the S-matrix are related as 
\begin{equation}
\mathbb{I}_{m}\widehat{G}\left( E\right) =\mathbb{I}_{m}\widehat{G}%
_{0}\left( E\right) +\frac{1}{4i}\widehat{S}^{-1}\left( E\right) \frac{%
\overleftrightarrow{\partial }}{\partial E}\widehat{S}\left( E\right)
\label{17}
\end{equation}%
where $\widehat{S}\left( E\right) $ is the scattering operator at the energy 
$E$ and $\widehat{G}_{0}\left( E\right) $ is the free part of the full
propagator (that leads to non-conected graphs in the cluster expansion). The
relation to the physical $T\left( E\right) $ matrix is 
\begin{equation}
\widehat{S}\left( E\right) =1+i\delta \left( E-\widehat{H}_{0}\right) 
\widehat{T}\left( E\right) ,  \label{18}
\end{equation}%
$\widehat{H}_{0}$ being the free Hamiltonian. The logarithm of the trace of $%
Z\left( T,V\right) $ can be written as [11] 
\begin{equation}
\ln Z\left( T,V\right) =\ln Z_{0}\left( T,V\right) +Tr\int e^{-\overline{b}%
.P}d^{4}P\left\{ \frac{-1}{\pi }\delta ^{3}\left( \overline{P}-\widehat{%
\overline{P}}\right) \mathbb{I}_{m}\left[ \widehat{G}\left( E\right) -%
\widehat{G}_{0}\left( E\right) \right] \right\} _{connected},  \label{19}
\end{equation}%
where the four-vector temperature $\overline{b}^{\mu }$ is defined by the
identity 
\begin{equation}
\overline{b}^{\mu }=\frac{1}{T}u^{\mu }  \label{20}
\end{equation}%
$T$ is the temperature in the rest frame of the box enclosing the
thermodynamical system, $u^{\mu }u_{\mu }=1\ $being its four-velocity. The
index $connected$ means that only the connected part of the full propagator $%
\widehat{G}\left( E\right) $ must be taken: simply by substracting to the
full propagator the free part $\widehat{G}_{0}\left( E\right) $
corresponding to an ideal gas configuration.

Recalling that the operation of taking the connected part leaves the
operations invariant, we can write 
\begin{equation}
\ln Z\left( T,V\right) =\ln Z_{0}\left( T,V\right) +\int e^{-\overline{b}%
.P}d^{4}P\ \rho _{I}\left( P^{2},\mathbb{V}.P,\mathbb{V}^{2}\right)
\label{21}
\end{equation}%
\begin{equation}
\mathbb{V}^{\mu }\mathbb{\equiv }\frac{2V}{\left( 2\pi \right) ^{3}}u^{\mu
}\ \ \text{(four-vector volume),}  \label{22}
\end{equation}%
where, by analogy with the interaction level density, the function which we
call mass spectrum, or cluster level density, is given by 
\begin{equation}
\rho _{I}\left( P^{2},V.P,V^{2}\right) =\left\{ \frac{-1}{\pi }\delta \left( 
\overline{P}-\widehat{\overline{P_{0}}}\right) \mathbb{I}_{m}\left[ \widehat{%
G}\left( E\right) -\widehat{G}_{0}\left( E\right) \right] \right\}
_{connected}  \label{23}
\end{equation}%
here the index $I$ means \textquotedblright interaction\textquotedblright .
On the other hand, from the pure statistical point of view, 
\begin{equation}
\rho _{I}\left( P^{2},V.P,V^{2}\right) =\rho -\rho _{0},  \label{24}
\end{equation}%
$\rho $ being the full density of states of the system under consideration
and $\rho _{0}$ the density of states in the free configuration, explicitly 
\begin{equation}
\rho _{0}=\overset{\infty }{\underset{k=1}{\sum }}\mathbb{V}^{\mu }P_{\mu
}\delta \left( P^{2}-k^{2}m^{2}\right)  \label{25}
\end{equation}%
$\rho $ satisfies the following relation: 
\begin{equation}
\rho \left( P^{2},V.P,V^{2}\right) =\mathbb{V}^{\mu }P_{\mu }\text{ }\rho
\left( \sqrt{P^{2}}\right)  \label{26}
\end{equation}%
The above condition is important. This means that the cluster counting $\rho
\left( P^{2},V.P,V^{2}\right) $ in the rest frame of the system can be
reexpressed by the counting of states for a single particle of the $\rho
\left( m\right) $ mass degeneracy moving in the volume $V$. Now, since the
mathematical mapping between the microcanonical and canonical ensembles is
easy to see, the $\Omega \left( E,V\right) $ we looking for is 
\begin{equation}
\ln \Omega \left( E,V\right) =\ln \Omega _{0}\left( E,V\right) +Tr\int
d^{4}P\left\{ \frac{-1}{\pi }\delta ^{3}\left( \overline{P}-\widehat{%
\overline{P}}\right) \mathbb{I}_{m}\left[ \widehat{G}\left( E\right) -%
\widehat{G}_{0}\left( E\right) \right] \right\} _{conn}  \label{27}
\end{equation}%
Notice that in order to obtain the above result, the main difference of our
procedure with the procedure given in refs.[12,18] for the canonical
ensemble is that we take as a starting point the propagator operator
eq.(16), with its statistical part. Considering that the imaginary part $%
\mathbb{I}_{m}\left[ \widehat{G}\left( E\right) -\widehat{G}_{0}\left(
E\right) \right] $ is the connected part of the full statistical propagator $%
\widehat{G}\left( E\right) ,$ it corresponds in the microcanonical field
formulation to the physical component $\Delta _{E}^{11}\left( k\right) $
eq.(14), which leads to an analog expression like (27), we can easily seen
from the comparison between the propagator, eq.(16), with expressions,
eq.(23-26), yielding the level density of states $\rho _{I\text{ }}$that the
interactions between particles in the system (dynamical information from the
connected part of the propagator (16)) are automatically translated into
variation of the energy ( mass ) levels in the statistical ensemble (given
by $\rho \left( m\right) $ and the statistical information of the propagator
(16)). That is, the study of any simple interactions between particles is
equivalent to that of the termodynamical properties of the statistical
ensemble of such particles as a whole.

\section{The Nambu-Goto action and the microcanonical propagator}

It is difficult to study this system in the Hamiltonian framework because of
the constraints and the vanishing of the Hamiltonian. As is known, the
Nambu-Goto action\ is invariant under the reparametrizations 
\begin{equation*}
\tau \rightarrow \widetilde{\tau }=f_{1}\left( \tau ,\sigma \right) \text{ }%
\ \ \sigma \rightarrow \widetilde{\sigma }=f_{2}\left( \tau ,\sigma \right)
\end{equation*}%
then, we can make the following choice for the dynamic variable $x_{0}$ and
the space variable $x_{1}$, as first proposed by B. Barbashov and N.
Chernikov in ref.[13] which do not restricts the essential physics and
simplifies considerably the dynamics of the system 
\begin{equation*}
x_{0}\left( \tau ,\sigma \right) \equiv x_{0}\left( \tau \right) ;\ \ \
x_{1}\left( \tau ,\sigma \right) \equiv \kappa \sigma \text{ \ \ \ \ }\left(
\kappa =const\right)
\end{equation*}%
for this, it is sufficient to use the chain rule of derivatives and to write
the action in the form 
\begin{equation*}
S=-\frac{\kappa }{\alpha ^{\prime }}\int_{\tau 1}^{\tau 2}\overset{.}{x}%
_{0}d\sigma \ d\tau \ \sqrt{\left[ 1-\left( \partial _{0}x_{b}\right) ^{2}%
\right] \left[ 1+\left( \partial _{1}x_{a}\right) ^{2}\right] }
\end{equation*}%
$\ \ \ \ \ \ \ \ $ $a,b=2,3;\ \partial _{1}x_{a}=\varepsilon _{1a}^{\ \ \
0b}\partial _{0}x_{b}$, where in order to simplify at maximum this action we
choose an orthonormal frame. (Thus, we pass from the Nambu-Goto action to
the Born-Infeld representation).

Therefore, the invariance with respect to the invariance of the coordinate
evolution parameter means that one of the dynamic variables of the theory ($%
x_{0}\left( \tau \right) $ in this case) becomes the observed time with the
corresponding non-zero Hamiltonian 
\begin{equation}
H_{BI}=\Pi _{a}\overset{.}{x}^{a}-L=\sqrt{\alpha ^{2}-\Pi _{b}\Pi ^{b}},
\label{28}
\end{equation}%
where 
\begin{equation*}
\Pi ^{b}=\frac{\partial L}{\partial \left( \partial _{0}x_{b}\right) }%
\hspace{0.5cm},\hspace{1cm}\alpha \equiv \frac{\kappa \sqrt{1+\left(
\partial _{1}x_{a}\right) ^{2}}}{\alpha ^{\prime }}
\end{equation*}%
Now in order to find the free propagator from the NG Hamiltonian we proceed
as follows:

From the simplest quantum path-integral formalism, we have 
\begin{equation*}
K\left( q^{\prime },t,q,0\right) \equiv \left\langle q^{\prime }\left\vert
\left( e^{H\varepsilon }\right) ^{N}\right\vert q\right\rangle =\left\langle
q^{\prime }\left\vert \Psi \left( r,s,t...\right) \right. \right\rangle ,
\end{equation*}%
where $K\left( q^{\prime },t,q,0\right) $ is the propagator, $H$ is the
Hamiltonian of the theory $t$ is the time that was fractionated in small
lapses $t=N\varepsilon $ and $q,q^{\prime },$ and $\Psi \left(
r,s,...\right) $ are the physical states with $r,s...$ quantum numbers. With
the usual path integral operations and introducing the integral
representation for a pseudodifferential operator [14] 
\begin{equation*}
\int \left( t^{2}+u^{2}\right) ^{-\lambda }e^{itx}dt=\frac{2\pi ^{1/2}}{%
\Gamma \left( \lambda \right) }\left( \frac{\left\vert x\right\vert }{2u}%
\right) ^{\lambda -1/2}K_{\lambda -1/2}\left( u\left\vert x\right\vert
\right)
\end{equation*}%
where $K_{\nu }\left( x\right) $ is the MacDonald's function, the propagator
for a sub-interval takes the form 
\begin{equation*}
K_{q_{j},q_{j+1}}=\delta _{q_{j},q_{j+1}}-i\varepsilon \left[ 4\alpha \frac{%
K_{-1}\left( \alpha \left\vert q_{j}-q_{j+1}\right\vert \right) }{\left\vert
q_{j}-q_{j+1}\right\vert }\right]
\end{equation*}

Putting on all the subinterval propagators together, we obtain the full
propagator 
\begin{equation*}
K=\delta _{q_{N},q_{0}}-iN\varepsilon \left[ 4\alpha \frac{K_{-1}\left(
\alpha \left| q_{N}-q_{0}\right| \right) }{\left| q_{N}-q_{0}\right| }\right]
\end{equation*}

Making, without loss of generality, the transformation $-it\rightarrow
-\beta $, $q_{0}=q_{N}$ integrating and Fourier transforming to momentum
space, yields the canonical partition function $Z_{c}$%
\begin{eqnarray*}
Z_{c} &=&\underset{N}{\sum }\left\{ 1-\beta \left[ 4\alpha \frac{%
K_{-1}\left( \alpha \left\vert q_{N}-q_{0}\right\vert \right) }{\left\vert
q_{N}-q_{0}\right\vert }\right] \right\} \\
&=&\underset{N}{\sum }\exp \left\{ -\beta \left[ 4\alpha \frac{K_{-1}\left(
\alpha \left\vert q_{N}-q_{0}\right\vert \right) }{\left\vert
q_{N}-q_{0}\right\vert }\right] \right\}
\end{eqnarray*}%
The microcanonical partition function $\Omega _{m}$ is obtained as the
inverse Laplace transform of the last expression : 
\begin{equation*}
\Omega _{m}=\delta \left( E\right) -\overset{\infty }{\underset{N=1}{\sum }}%
\underset{\overline{p}_{1}}{\sum }\overset{\infty }{\underset{n_{1}=1}{\sum }%
}.......\underset{\overline{p}_{N}}{\sum }\overset{\infty }{\underset{n_{n}=1%
}{\sum }}\left[ 4\alpha \frac{K_{-1}\left( \alpha \left\vert \sum
n_{j}\varepsilon _{j}-E\right\vert \right) }{E^{2}}\frac{1}{%
n_{1}n_{2}.....n_{N}}\right] ,
\end{equation*}%
where the factor $\frac{1}{n_{1}n_{2}.....n_{N}}$ allows one to eliminate
the overcounting, and $\sum n_{j}\varepsilon _{j}=E_{N}$.

The free microcanonical propagator for the N-extended body system can be
consistently formulated using the relation (valid for free fields) between
time ordered products and normal products 
\begin{equation*}
-iT\left[ \varphi \left( x\right) \varphi \left( 0\right) \right]
=D_{F}\left( x\right) -i:\varphi \left( x\right) \varphi \left( 0\right) :
\end{equation*}%
where $T$ is the temperature of the thermal bath, $D_{F}\left( x\right)
=-i\left\langle \mathcal{J}\left\vert \varphi \left( x\right) \varphi \left(
0\right) \right\vert \mathcal{J}\right\rangle $ is the ordinary Feynman
propagator with the expectation value evaluated in the basic states of our
system ( i.e., for zero temperature ) 
\begin{equation*}
\left\vert \mathcal{J}\right\rangle =\left[ \underset{k,m}{\prod }\underset{%
n_{k},m}{\sum }\right] \underset{k,m}{\prod }\left\vert n_{k,m}\right\rangle
\otimes \left\vert \widetilde{n}_{k,m}\right\rangle
\end{equation*}%
Since the relation between microcanonical-canonical formulations is via a
Laplace transform, it is reasonable to perform the following mapping: 
\begin{equation*}
\int_{0}^{\infty }dE^{\prime }\Omega _{E-E^{\prime }}D_{E^{\prime }}=
\end{equation*}%
\begin{equation}
=D_{F}\Omega _{E}-i\overset{\infty }{\underset{N=1}{\sum }}\underset{%
\overline{p}_{1}}{\sum }\overset{\infty }{\underset{n_{1}=1}{\sum }}.......%
\underset{\overline{p}_{N}}{\sum }\overset{\infty }{\underset{n_{n}=1}{\sum }%
}\left[ 4\alpha \frac{K_{-1}\left( \alpha \left\vert \sum n_{j}\varepsilon
_{j}-E\right\vert \right) }{E^{2}}\frac{1}{n_{1}n_{2}.....n_{N}}\right]
\left\langle \mathcal{J}\right\vert :\varphi \left( x\right) \varphi \left(
0\right) :\left\vert E\right\rangle ,  \label{29}
\end{equation}%
where we defined the microcanonical state as 
\begin{equation*}
\left\vert E\right\rangle =\frac{1}{\Omega _{E}}\int_{0}^{\infty }dE^{\prime
}\Omega _{E-E^{\prime }}L_{E^{\prime }}^{-1}\left[ \left\vert \beta
\right\rangle \right] ,
\end{equation*}%
L$^{-1}$ being the inverse Laplace transform and $D_{E^{\prime }}$ being the
microcanonical propagator. The matrix element for the most general states in
our system is 
\begin{equation*}
\left\langle \mathcal{J}\right\vert :\varphi \left( x\right) \varphi \left(
0\right) :\left\vert E\right\rangle =\frac{1}{\Omega _{E}}\Omega
_{E-E^{\prime }}\underset{\overline{p}}{\sum }\left[ \frac{n}{\varepsilon
_{p}V}\cos \left( \overline{p}.\overline{x}-\varepsilon _{p}t\right) \right]
\end{equation*}%
By inserting this expression in the definition of the microcanonical
propagator $D_{E^{\prime }}$ (39) given above and converting the momentum
sum into an integral, we have 
\begin{equation*}
D_{E}(t,\overline{x})=\delta \left( E\right) D_{F}-4i\alpha \int \frac{d^{3}p%
}{\left( 2\pi \right) ^{3}\varepsilon _{j}}\overset{}{\underset{n=1}{\overset%
{\infty }{\sum }}}\frac{K_{-1}\left( \alpha \left\vert n_{j}\varepsilon
_{j}-E\right\vert \right) }{E^{2}}\frac{\Omega \left( E-n_{j}\varepsilon
_{j}\right) }{\Omega \left( E\right) }\cos \left( \overline{p}.\overline{x}%
-\varepsilon _{j}t\right)
\end{equation*}%
Finally, Fourier transform to momentum representation gives 
\begin{equation*}
D_{E}(t,\overline{x})=\frac{\delta \left( E\right) }{\omega
^{2}-k^{2}-m^{2}+i\varepsilon }-
\end{equation*}%
\begin{equation}
-8\pi i\alpha \delta \left( \omega ^{2}-k^{2}-m^{2}\right) \overset{}{%
\underset{l=1}{\overset{\infty }{\sum }}}\frac{K_{-1}\left( \alpha
\left\vert l\omega _{k}-E\right\vert \right) }{E^{2}}\frac{\Omega \left(
E-l\omega _{k}\right) }{\Omega \left( E\right) }\theta \left( E-l\omega
_{k}\right) ,  \label{30}
\end{equation}%
where $\theta \left( x\right) $ is the usual step function. The first term
in the microcanonical propagator is the usual (non-termal) Feynman
propagator, the second one is the new microcanonical statistical part. This
part is crucial for the correct description of the full N-extended body
system, as we can see explicitly expanding the Mac Donald's function $K_{-1}$
in the second term of the free microcanonical propagator 
\begin{equation*}
\begin{array}{lll}
& 8\pi i\alpha \delta \left( \omega ^{2}-k^{2}-m^{2}\right) \underset{l=1}{%
\overset{M/\omega _{k}}{\sum }}\left[ \ln \left( \frac{\gamma _{e}\alpha
\left\vert l\omega _{k}-E\right\vert }{2}\right) \frac{\alpha \left\vert
l\omega _{k}-E\right\vert }{2}\underset{s=0}{\overset{\infty }{\sum }}\frac{%
\left( \alpha \left\vert l\omega _{k}-E\right\vert /2\right) ^{2s}}{s\Gamma
\left( s+2\right) }-\right. &  \\ 
&  &  \\ 
& \left. \frac{1}{2}\underset{s=0}{\overset{\infty }{\sum }}\frac{\left(
\alpha \left\vert l\omega _{k}-E\right\vert /2\right) ^{2s+1}}{s!\Gamma
\left( s\right) }\left( \underset{h=1}{\overset{s}{\sum }}\frac{1}{h}+%
\underset{h=1}{\overset{s+1}{\sum }}\frac{1}{h}\right) +\frac{\alpha
\left\vert l\omega _{k}-E\right\vert }{4}\right] \frac{\theta \left(
E-l\omega _{k}\right) }{E^{2}}\frac{\Omega \left( E-l\omega _{k}\right) }{%
\Omega \left( E\right) } & 
\end{array}%
\end{equation*}%
and expressing it as a function of the mass$,\left( M\right) $ being the
energy E of the extended-bodies (strings) in our system ($E=M\equiv $ total
mass of the system ) 
\begin{equation*}
\begin{array}{lll}
& 8\pi i\alpha \delta \left( \omega ^{2}-k^{2}-m^{2}\right) \underset{l=1}{%
\overset{M/\omega _{k}}{\sum }}\left[ \ln \left( \frac{\gamma _{e}\alpha
\left\vert l\omega _{k}-M\right\vert }{2}\right) \frac{\alpha \left\vert
l\omega _{k}-M\right\vert }{2}\underset{s=0}{\overset{\infty }{\sum }}\frac{%
\left( \alpha \left\vert l\omega _{k}-M\right\vert /2\right) ^{2s}}{s\Gamma
\left( s+2\right) }-\right. &  \\ 
&  &  \\ 
& \left. \frac{1}{2}\underset{s=0}{\overset{\infty }{\sum }}\frac{\left(
\alpha \left\vert l\omega _{k}-M\right\vert /2\right) ^{2s+1}}{s!\Gamma
\left( s\right) }\left( \underset{h=1}{\overset{s}{\sum }}\frac{1}{h}+%
\underset{h=1}{\overset{s+1}{\sum }}\frac{1}{h}\right) +\frac{\alpha
\left\vert l\omega _{k}-M\right\vert }{4}\right] \frac{\theta \left(
M-l\omega _{k}\right) }{M^{2}}\frac{\Omega \left( M-l\omega _{k}\right) }{%
\Omega \left( M\right) } & 
\end{array}%
\end{equation*}%
We see that when M$\rightarrow 0$ this expression yields the pure
string-like behaviour (Gamma type string-amplitude). Notice that previously
[15,16] the relation between the Feynman propagator and the Veneziano
amplitude was put \textquotedblright by hand\textquotedblright . We see that
this type of structure coming from the statistical microcanonical part is
contained in our microcanonical propagator, eq.(30). It should be noted that
the relation between temporal and normal ordering of the field operators
contributes to the statistical part of this full propagator. These
implications will be discussed everywhere [17], where we will focus on some
concrete problems (arrow of time, the early universe, etc.).

It is interesting to note that the main difference between our
microcanonical propagator, eq.(30), and the $\Delta _{11}$ of the previous
section is in that the propagator, eq.(30), includes all nonlocal effects:
from the n-bodies of the system as extended objects (i.e., strings) and from
the derivation of this propagator from a theory with a Hamiltonian not
quadratic in momenta as in the Nambu-Goto formulation of the string theory.
For instance, the propagator, eq.(30), becomes the propagator, eq.(14), when
the extended bodies (e.g., strings) became point-particles and the
Hamiltonian is quadratic in momenta (constrained particle Hamiltonian).

\section{Acknowledgments}

I am very grateful to professors A. Dorokhov and N.G. Sanchez for very
useful discussions and suggestions and to Galina Sandukovskaya for help me
with the english grammar of this paper. Thanks also are due to the people of
the Bogoliubov Laboratory of Theoretical Physics and Directorate of the
JINR\ for their hospitality and support.

\section{References}

[1]. C. Bernard, Phys. Rev. \textbf{D 9}, 3312 (1974).

[2]. L. Dolan and R. Jackiw, Phys. Rev. \textbf{D 9}, 3320 (1974).

[3]. S. Weinberg, Phys. Rev. \textbf{D 9}, 3357 (1974).

[4]. T. Matsubara, Prog. Theor. Phys. \textbf{14}, 351 (1975).

[5]. A. J. Niemi and G. W. Semenoff, Ann. Phys. (N.Y.) \textbf{152}, 105
(1984); Nucl. Phys. \textbf{B 230,}[FS10] 181 (1984).

[6]. Y. Takahashi and H. Umezawa, Collective Phenomena \textbf{2}, 55 (1975).

[7]. N. P. Landsman and Ch. G. Van Weert, Phys. Rep. \textbf{145}, 141
(1987).

[8]. H. A. Weldon, Ann. Phys. (N.Y.) \textbf{193}, 166 (1989).

[9]. H. Umezawa, H. Matsumoto and M. Tachiki, \textit{Thermo Field Dynamics
and Condensed States, }North Holland Publishing Co.,Amsterdam (1982).

[10]. R. Casadio and B. Harms, Phys. Rev. \textbf{D58}, 044014 (1998).

[11]. R. Dashen et al., Phys. Rev.\textbf{187}, 345 (1969).

[12]. L. Sertorio and M. Toller, N.C., \textbf{14 A}, 21 (1973).

[13] B. M. Barbashov and N. A. Chernikov, Zh. Eksp. Theor. Fiz. (JETP) 
\textbf{50}, 1296 (1966); \textbf{51}, 658 (1966); Comm. Math. Phys. \textbf{%
5}, 313 (1966).

[14]. Yu. A. Brichkov and A. P. Prudnikov, \textit{Integral transform of
General Functions, }Nauka, Moscow (1977).(In Russian).

[15]. G. Veneziano, Phys. Rept. \textbf{9}, 199 (1974).

[16]. J. Scherk, Rev. Mod. Phys. \textbf{47}, 123 (1975).

[17] D. J. Cirilo-Lombardo et al., work in preparation.

[18]. R. Dashen, S. Ma and H. J. Bernstein, Phys. Rev. \textbf{D 187}, 345
(1969).

[19]. L. L. Jenkovszky, A. A. Trushevsky and L. Sertorio, Lett. Nuov. Cim. 
\textbf{15}, 200 (1976).

\end{document}